# The most energetic cosmic rays and their possible sources

L.G. Dedenko[a], D.A. Podgrudkov[a], T.M. Roganova[b] and G.F. Fedorova[b]

[a] Physical Department of Moscow State University, Moscow 119992, Leninskie Gory, Russia
[b] Skobeltsyn Institute of Nuclear Physics of Moscow State University, Moscow 119992, Leninskie Gory, Russia
 e-mail: ddn@dec1.sinp.msu.ru

## Abstract

The primary cosmic rays particles with energies above $10^{20}$ eV have been observed at many extensive air shower arrays since the beginning of observations over 40 years ago. The validity of measurements of signal $s(600)$ used as energy estimation parameter at the Yakutsk array has been confirmed. Our calculations show that the width of the time pulses increases from nearly 100 ns at a distance of 100 m from the shower axis up to 4 − 5 μs at 1500 m. The calculated estimate of energy of extensive air shower is ~ 1.7 times smaller than the experimental estimate for the same value $s(600)$. The pointing directions of extensive air showers observed at the Pierre Auger Observatory were fitted within ±3.1° with positions of the nearby active galactic nuclei from the Veron-Cetty and P. Veron catalog. The cosmic ray luminosity of the active galactic nuclei which happened to be a source of the particular cosmic ray event constitutes a fraction ~$10^{-4}$ of the optical one if only cosmic ray particles with energies above $6·10^{19}$ eV are produced. If produced cosmic ray particles have a spectrum up to ~ 100 GeV then the cosmic ray luminosity of the active galactic nuclei should be much higher than the optical one.

## Introduction

It is remarkable that the extensive air showers (EAS) with energies above $10^{20}$ eV is already more than 40 years. First observation was made by J. Linsley at the beginning of 60-th [1] of last century at the Volcano Ranch array. Then followed observations at the Yakutsk array (YA) [2], the Akeno Giant Air Shower Array (AGASA) [3] and Fly's Eye array [4]. Finally, the High Resolution Fly's Eye Collaboration (HiRes) [5] claimed the observation of giant air shower with energy $1.1·10^{20}$ eV and the Pierre Auger Observatory Collaboration (PAO) announced the existence of a shower with energy above $1.4·10^{20}$ eV [6]. The Greizen-Zatsepin-Kuzmin (GZK) effect [7, 8] have been claimed to be observed both by the HiRes [9] and PAO [10]. However, some nearby sources of cosmic rays might produce particles with energies above $10^{20}$ eV.

There are different methods of energy estimation of EAS. The total number $N$ of particles for the vertical shower inferred from the number $N(\Theta)$ for inclined shower with the zenith angle $\Theta$ with the help of measured longitudinal attenuation of showers may be used as energy estimator. Some parameters of the lateral distribution function (LDF) measured by surface detectors (in combination with model calculations for vertical showers and with additional methods for inclined showers, such as constant intensity cuts method (CICM) [11] are often used. For calibration of ground detectors one may use the Vavilov-Cherenkov light or fluorescence light measured by optical detectors. The energy of EAS may be estimated with help of such parameters of LDF as signal $s(600)$ (or $s(1000)$), the density of deposited energy in the scintillation detector or a water tank at the distance of 600 m (or 1000 m) from the shower axis. Nevertheless, we suggest that all detector readings should be taken into account, not only one parameter, when estimating energy of individual EAS. Further, we suggest that energy estimations of the inclined showers should be carried out without CICM to suppress additional artificial errors. In addition, time distribution of signals in detectors should be compares with predictions (in terms of some model) after estimating LDF curve. So the time pulses of these signals should be calculated to be compared with the experimental data.

It is a known problem of the calibration of energy estimates of giant air showers. At the Yakutsk array the Vavilov-Cherenkov light is used to calibrate signals in scintillation detectors [12]. It is worth noting that our calculations [13] show that the same flux of the Vavilov-Cherenkov light should be accompanied by much higher signals in scintillation detectors than it is measured in the experiment [14].

This contradiction may have different explanations. First, the hadron interaction model does not correctly describe the real processes at high energies. This question will be probed at the large hadron collider (LHC) once it is operational. Second, we may have different chemical composition of primary cosmic rays − all our calculations were undertaken in assumption of light chemical composition at high energies, e.g. primary particles are protons only. And almost all recent experiments claim that primary particles at that energies are protons [9, 16]. Then, another possibility is in absolute calibration of scintillation detectors at the Yakutsk array with help of the Vavilov-Cherenkov radiation. This question should be clarified in the experiment. In addition, there is a possibility that signals are not measured correctly namely measured only partially due to short time gate.

Each detector array encounter the problem of full signal registration. Each detector after being triggered by some event collects signal during some period of time, a so-called time gate. Time gates should be wide enough to collect possibly all particles coming to the detector. On the other hand, while we always have a background from low energy cosmic rays or the Earth's sources (radioactivity, light pollution, etc.), the time gates should not be too wide to keep signal to noise ratio high enough. So, the time gates should be about of disk thickness. At the most detector arrays such as at the Haverah Park [17], Volcano Ranch, Yakutsk time gates were set about of 2 μs.

A. Watson [17] suggested that time gates at the Yakutsk array are too narrow to collect all particles. Thus the signals in scintillation detectors are underestimated, which leads to an overestimation of the primary particle energy. This is supported by a rapid change in the steepness of measured signal lateral distribution function along with the energy increase. In addition, some of showers detected at the Haverah Park had disk thickness more than 2.2 μs.

At the very beginning of measurements in the southern semi-sphere the Pierre Auger Collaboration has reported that arrival directions of ultrahigh energy cosmic rays (UHECR) are not isotropic [18, 19, 20]. Moreover a remarkable correlations between the pointing directions of the PAO events and positions of relatively nearby active galactic nuclei (AGN) from the Veron-Cetty and P. Veron (VCV) catalog [21] has been observed [18, 19, 20]. This profound discovery opens the new era of the cosmic ray astronomy [22, 23]. It is a puzzle that in the northern semi-sphere such correlations between pointing directions of UHECR and positions of the AGNs from the VCV catalog have not been observed [24].

Our analysis of data of the Yakutsk array [25] which is also in the northern semi-sphere showed rather isotropic distribution of arrival directions of ultrahigh energy cosmic rays than a clustering in the super galactic plane. But later analysis of the Yakutsk data [26] has claimed the correlation with the AGN.

Searching for correlations between the pointing directions of UHECR and positions of some prominent objects in the sky allows to unveil the possible sources of these cosmic rays. Such correlations with the quasars [27], some objects in the supergalactic plane [28], the BL Lacertae [29] and Seifert galaxies [30] have been reported.

A suggestion [18, 19, 20] that the nearby active galactic nuclei (or other objects with the same spatial distribution) may be the possible sources of UHECR (the AGN hypothesis) is very attractive. However this suggestion as it was pointed out in refs [21, 22] has some problems. Namely, a profound deficit of UHECR events from the Virgo cluster was remarked [31, 32]. It should be remembered that one of the paper has such subtitle as "All roads lead back to Virgo" [33].

In this paper we present some results of calculations of the time structure of signals in scintillation detectors of the Yakutsk array. In addition, we discuss some common requirements to the celestial objects, which are supposed to be the sources of the UHECR. Namely, we estimate the cosmic ray luminosity of the nearby AGN, which should be powerful enough to produce the PAO events.

## Time-dimensional structure of signals in the scintillation detectors of EAS

To clarify this time problem we made simulations of signal time distributions. For shower simulation we used modified CORSIKA 6.500 [34] and EGS4 [35] codes incorporating QGSJet-II model [36,37,38,39,40] for high energy hadron interactions and Gheisha-2002d [42] model for low energy hadron interactions. CORSIKA code was modified in order to printout all particles reaching the observation level. The following parameters were printed out: particle type, distance from the shower axis, particle energy, cosine of the incidence angle, time delay and particle weight. For simulation all parameters of atmosphere, magnetic field and observation level was set to fit the conditions of the Yakutsk array. Only vertical showers from primary protons were simulated. The time delay is time between arrival on observation level the first particle and current particle. Thinning procedure [42] parameter $\varepsilon$ was set equal to $10^{-6}$ eV.

To determine signals from EAS particles in scintillation detector the database of signals from gammas, electrons, positrons and muons of different energies and various zenith angles was created with GEANT4 code. The database included signals from particle with all possible incidence angles and different energies (from 1 MeV up to 10 GeV for gammas, electrons and positrons and from 0.3 GeV up to 1000 GeV for muons). For each particle in CORSIKA, output file corresponding signal (calculated with 5-point interpolation method) was assigned. Signals from particles in VEM/m$^2$ units (VEM - vertical equivalent muon that induces in scintillation detector ionization ~10.5 MeV) were summarized in bins by 1, 50 and 250 ns at 100, 600 and 1000 m and larger distances from shower axis correspondingly. The results of these calculations are presented at figures 1 − 5. Figure 1 presents the calculated signal impulse in scintillation detector of the Yakutsk array at 100 m distance from axis of $10^{18}$ eV proton induced shower. Signal is integrated within 1 ns bins. Figure 2 shows the calculated signal impulse at 600 m from the shower (same shower as at previous figure) axis. Signal is integrated within 50 ns bins. On figure 3 the signal impulse in detector at 1000 m distance from shower axis is presented (250 ns bins). Figure 4 shows the calculated signal at 1250 m distance from shower axis (solid line) with data from AGASA experiment [43] (dots with error bars) for the $2\cdot10^{18}$ eV proton induced shower. Signal is integrated within the 100 ns bins. On figure 5 the calculated time structure of signal at 1500 m from shower axis (primary particle $10^{18}$ eV proton) is presented. Arrival time of the first particle in selected area (observation level was divided into rings with 50 m width) was taken as initial moment for signal for analysis. From figures, one can see that at small distances from shower axis all signal is collected in 2 μs. At large distances from shower axis signal width exceeds 2 μs and considerable part of signal is not registered. Thus measured in experiment signals lateral distribution function (LDF) will differ from real signals LDF and difference between them at large distances from shower axis can be significant.

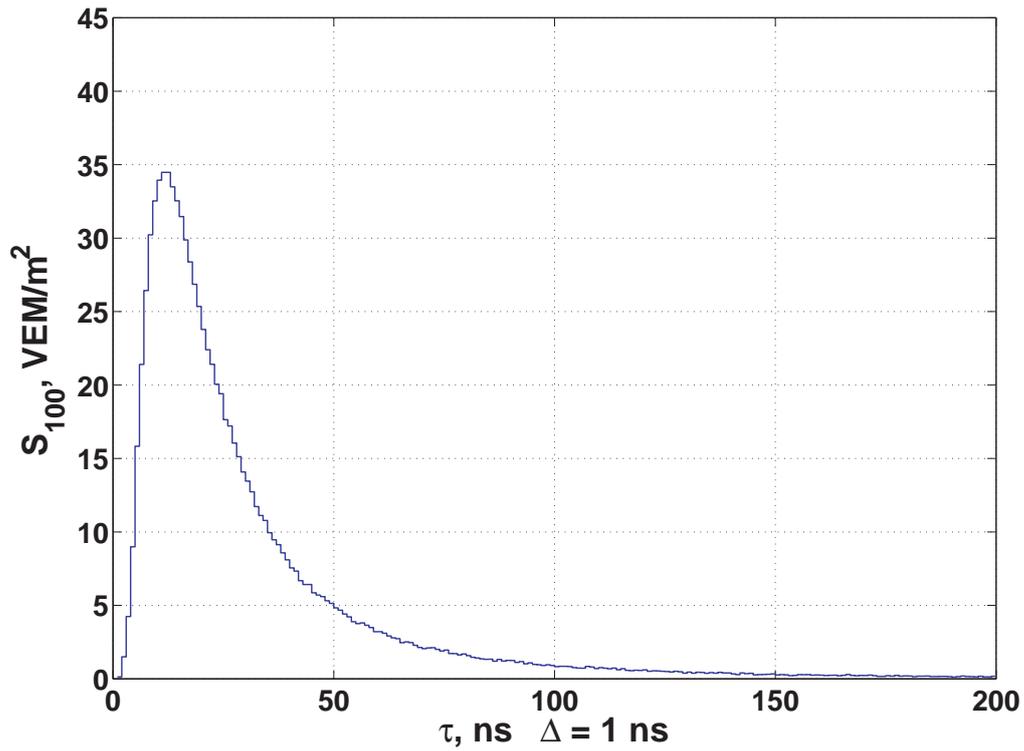

Fig. 1. Signal time-structure at the 100 m distance from the shower axis induced by primary proton with energy $10^{18}$ eV. Signal is summed within 1 ns bins.

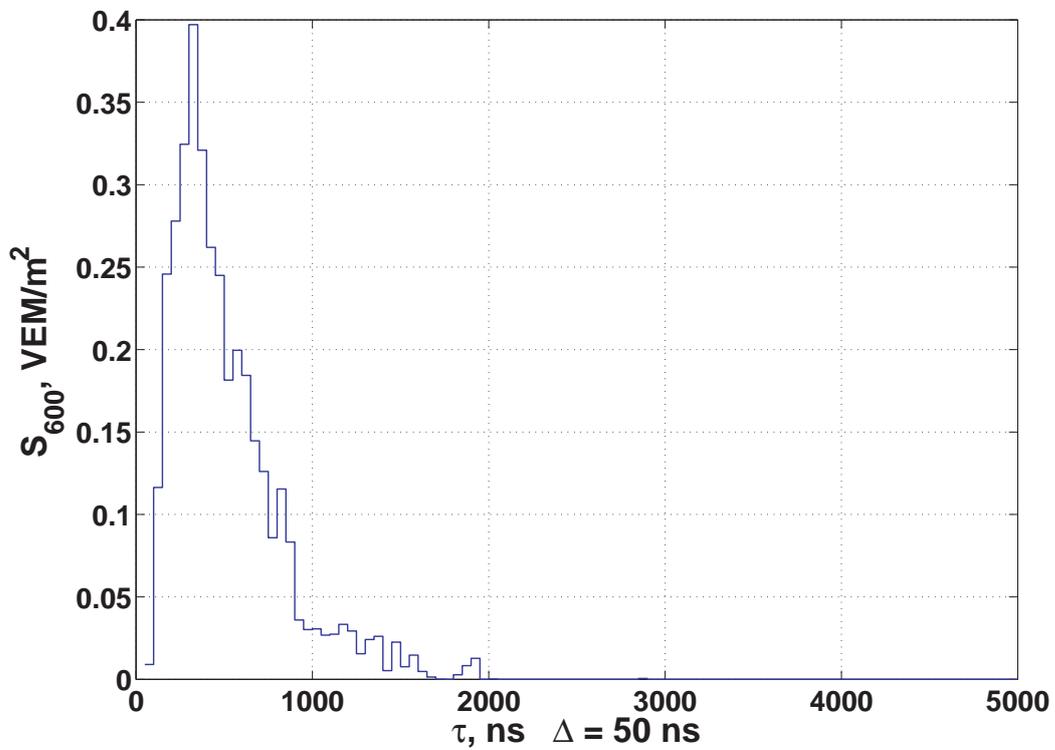

Fig. 2. Signal time-structure at the 600 m distance from the shower axis induced by primary proton with energy $10^{18}$ eV. Signal is summed within 50 ns bins.

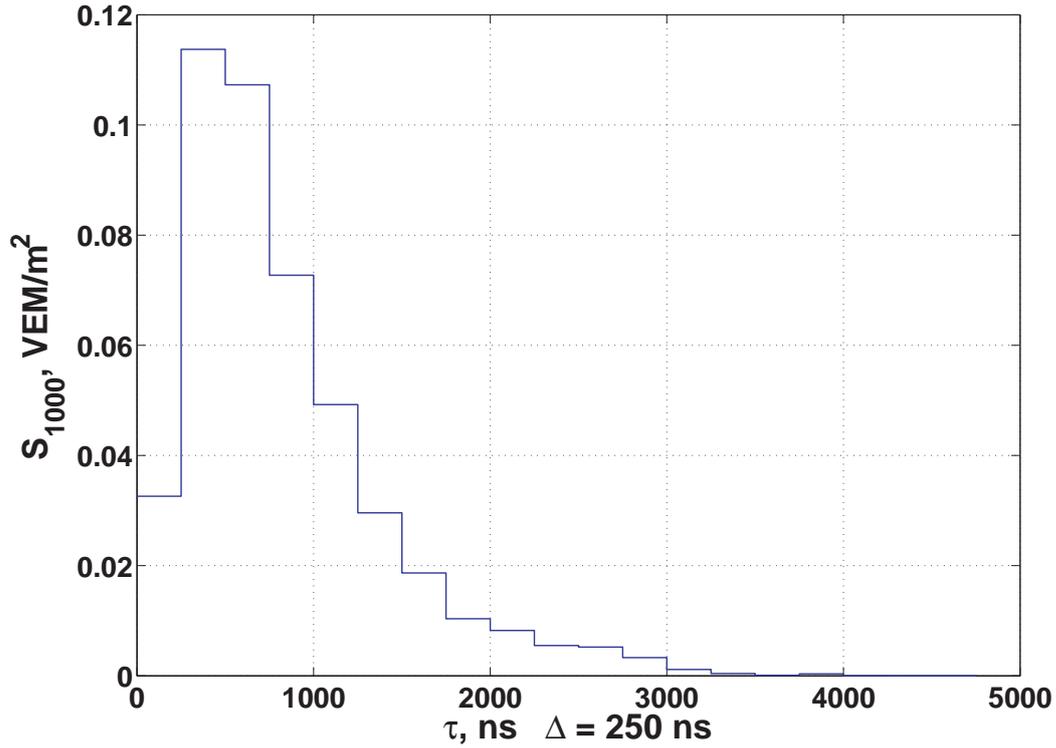

Fig. 3. Signal time-structure at the 1000 m distance from the shower axis induced by primary proton with energy $10^{18}$ eV. Signal is summed within 250 ns bins.

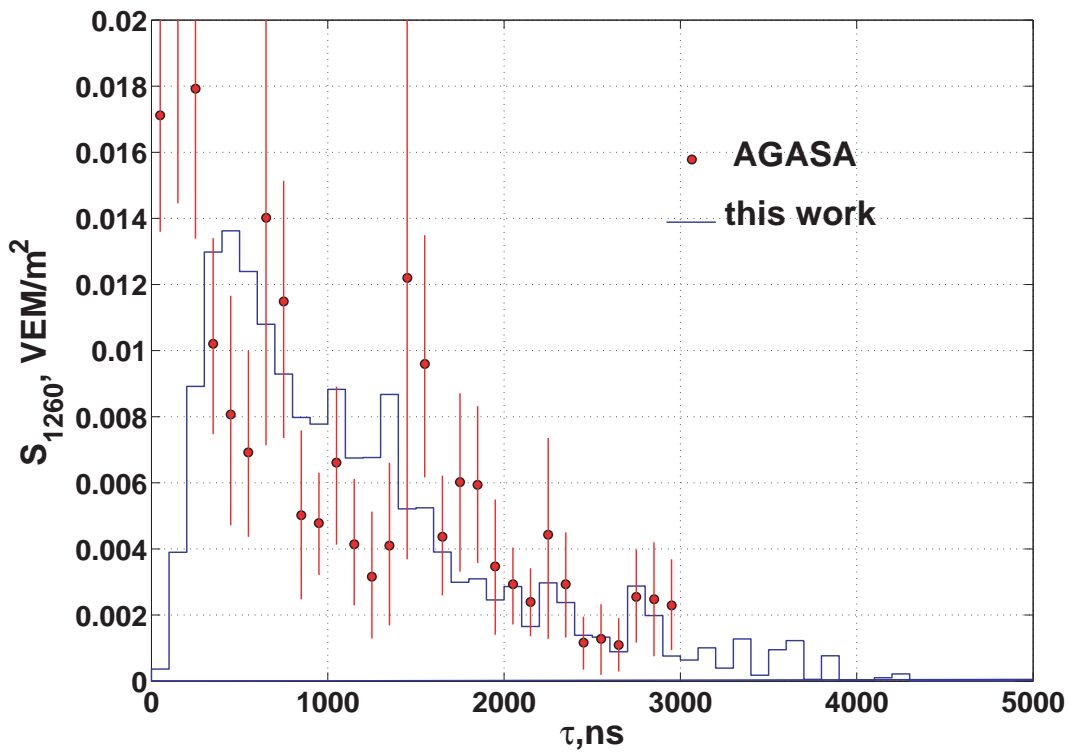

Fig. 4. Signal time-structure at the 1250 m distance from the shower axis induced by primary proton with energy $2 \cdot 10^{18}$ eV (solid curve). Signal is summed within 100 ns bins. Dots with errors present AGASA data [43].

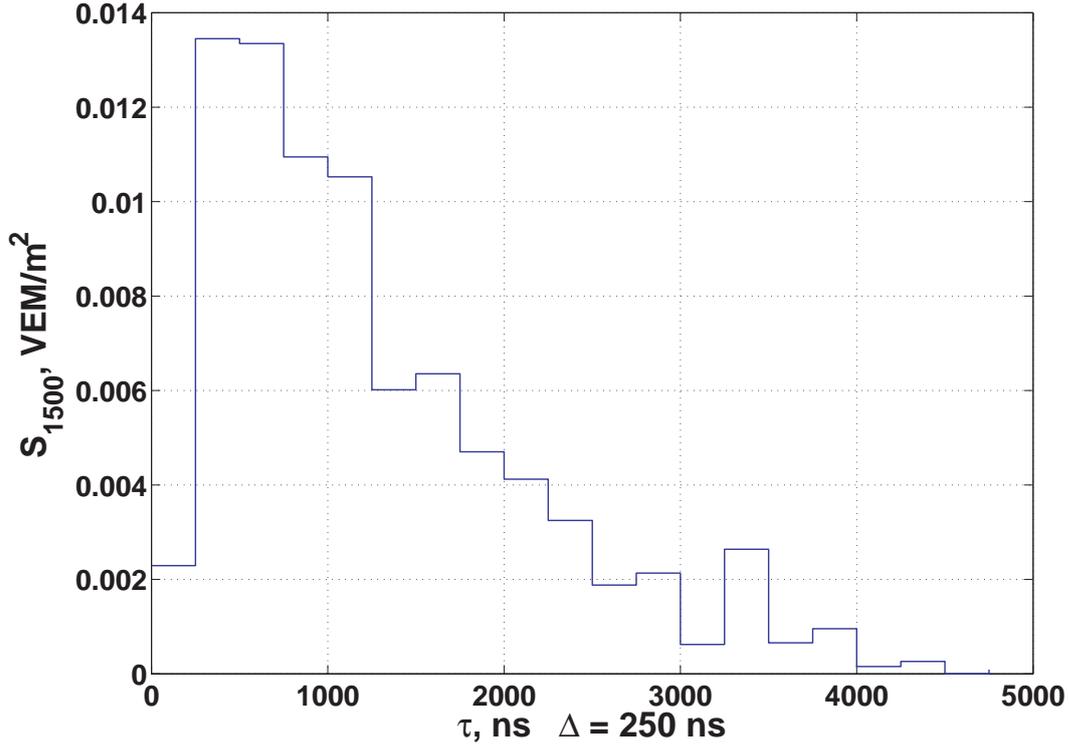

Fig. 5. Signal time-structure at the 1500 m distance from the shower axis induced by primary proton with energy $10^{18}$ eV. Signal is summed within 250 ns bins.

Figure 6 presents LDF of signals for showers with energies $10^{17} - 10^{21}$ eV, calculated with and without respect to signal collection time. At the distances about 1500 - 2000 m difference between them reaches 30 - 40% (see fig. 7). From this point one can make conclusion that main energy estimation parameter at Yakutsk array $s(600)$ − signal at 600 m distance from shower axis − is being measured in experiment correctly. Nevertheless, only in a relatively few events signal $s(600)$ (or close one) is measured directly, so it is reconstructed from signals LDF that in turn is reconstructed with data from large distances from the shower axis. Along with this possible difference between measured LDF and theoretical LDF requires further analysis.

We have obtained theoretical dependence of signal $s(600)$ on the primary proton energy. This data yielded to the following EAS energy estimation formula:

$$E_0 = 2{,}8 \cdot 10^{17} \, (s(600))^{0.99}, \qquad (1)$$

here $E_0$ is in eV, $s(600)$ is in VEM/m$^2$. This estimate differs from the one used at the Yakutsk array [14]:

$$E_0 = 4{,}8 \cdot 10^{17} \, (s(600))^{0.98} \qquad (2)$$

by about 1.7 times. New energy estimation lowers intensity of primary cosmic rays flux at Yakutsk array and brings it to an agreement with results of HiRes [45] and PAO [44].

Also on this data we performed analysis of shower front form and dependence of disk thickness on distance from shower axis. First particles hitting detector at selected distance from shower axis were considered as shower front. On figure 8 calculated forefront is shown (dots) with 2 approximations – spherical front and power function dependence of time lag versus distance from shower axis. One can see that shower forefront is better approximated by power function $\sigma_f = aR^b$ ($a = 1.47 \cdot 10^{-5}$, $b = 2.195$, $\chi^2 = 3{,}17$ per one degree of freedom) that by spherical front with radius $R = 17$ km ($\chi^2 = 110$ per one degree of freedom).

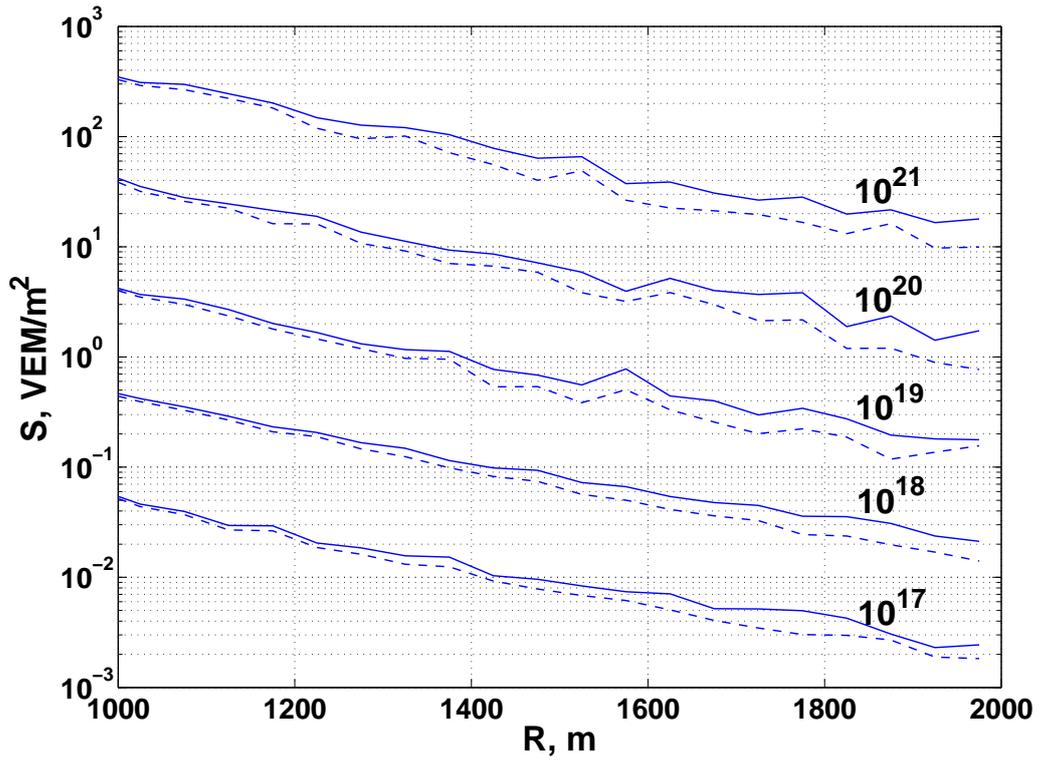

Fig. 6. Lateral distribution functions of signals from showers from primary protons with energies $10^{17} - 10^{21}$ eV calculated assuming infinite time-gates (dashed lines) and 2 μs width time gates (solid lines).

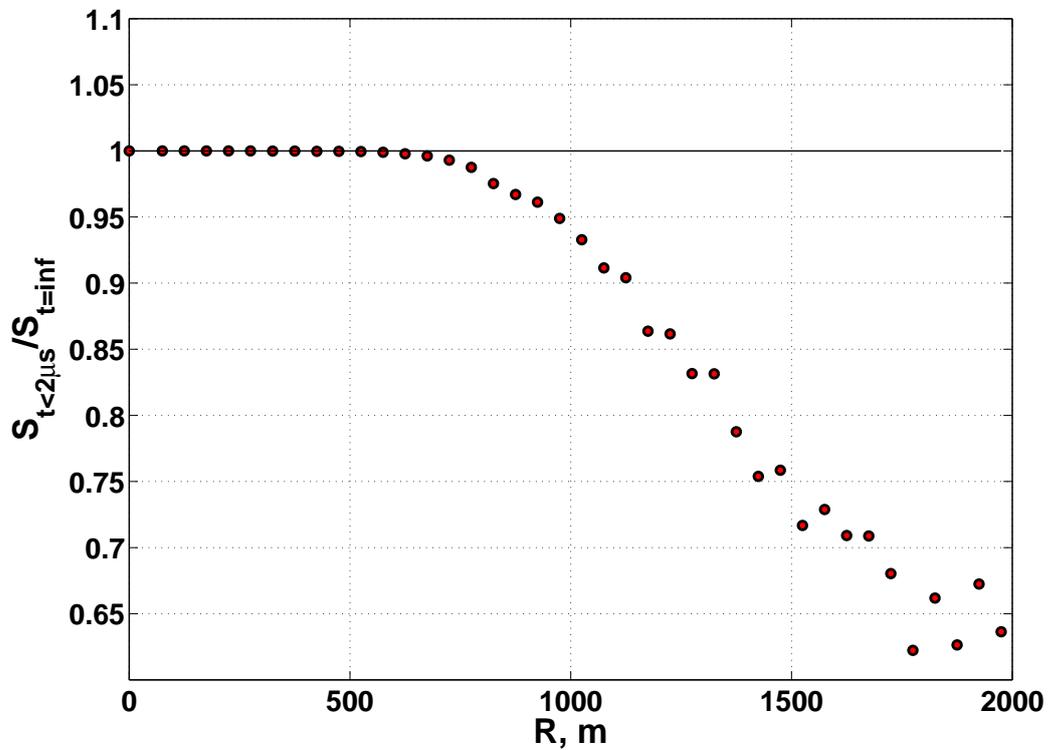

Fig. 7. The ratio of signal collected within 2 μs to the full signal at given distance from shower axis.

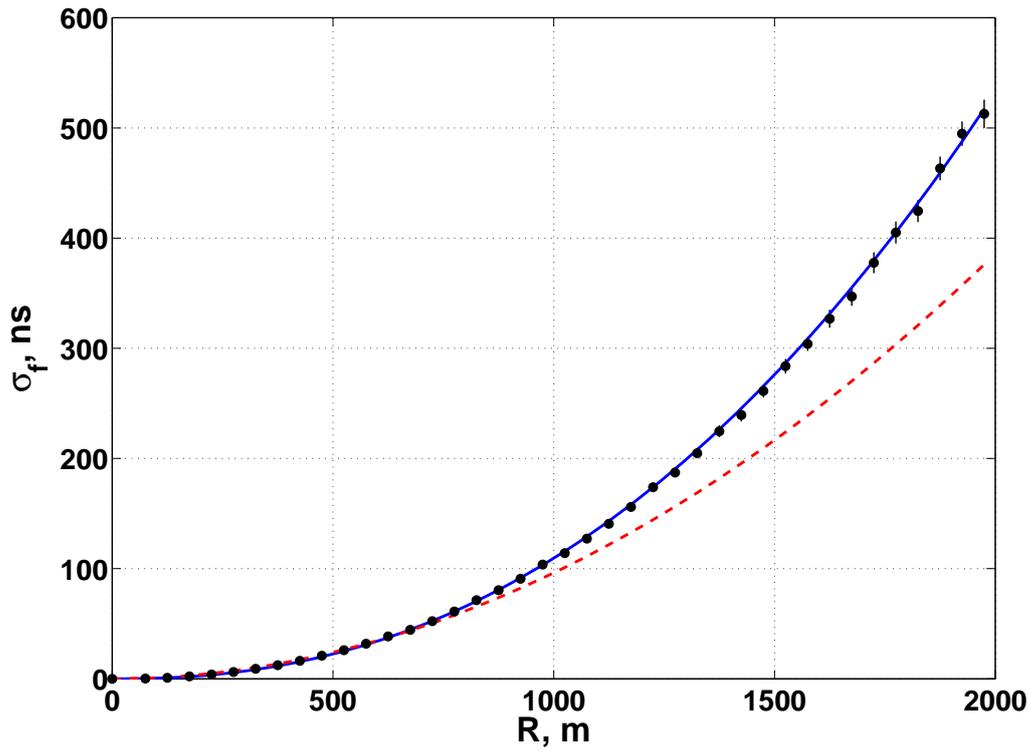

Fig. 8. Vertical shower forefront. Dots - calculated data, solid line − power function approximation (see in text), dashed line - spherical front best fit ($R = 17$ km).

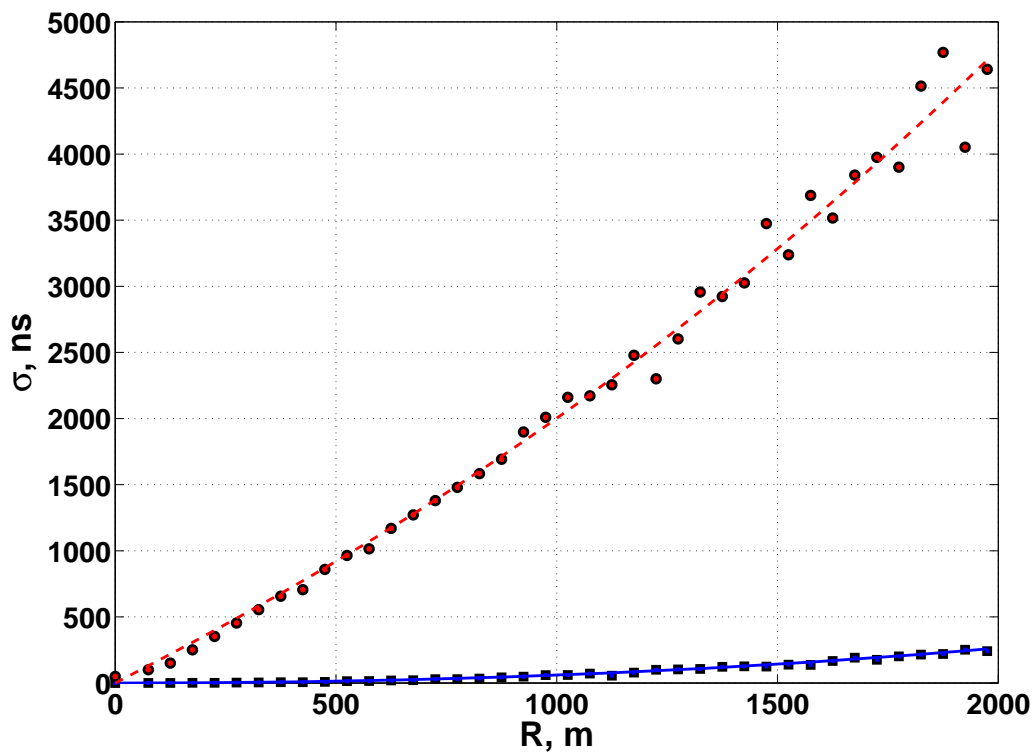

Fig. 9. Shower fore (squares) and back fronts (dots) and their approximations by power functions (see in text) - solid and dashed lines correspondingly.

In our calculations we considered disk thickness as time required to obtain 95% of signal at selected distance from shower axis. The results of this calculations are presented at figure 9. Note that the AGASA data shows $2.5 - 3$ μs disk thickness at distance 1260 m [43, 45] and from fig. 9 we obtain that disk thickness is about 2.7 μs. Shower back front was approximated by function $\sigma_b = aR^b + kR + c$. ($a = 3{,}98 \cdot 10^{-5}$, $b = 2{,}28$, $k = 1{,}73$, $c = 0$). One can see that at distances further than 1000 m disk thickness exceeds 2 μs.

## Cosmic ray luminosity of the nearby active galactic nuvlei

First, we have been searching for coincidences (within ±3.1°) of the pointing directions of the ultra high energy cosmic rays (UHECR) observed [20] with the positions of the AGN from the VCV catalog [21]. If a pointing direction of a particular EAS coincides (within ±3.1°) with positions of several AGN (e.g. with m objects) then each AGN (of this sample of m objects) was assigned a weight $w$ ($w=1/m$). To estimate the cosmic ray luminosity we assumed that any particular sources are viewed only a fraction 1/3 of the total exposure time. We ignore also the absorption of the ultrahigh energy cosmic rays due to the GZK effect [7, 8].

If the energy of 10 J is assigned to each UHECR particle and the observed intensity of such particles is assumed then in case when sources emit isotropic cosmic rays the cosmic ray luminosity $LCR$ of any object at distance $R$ from the Earth may be estimated by simple formula:

$$LCR_{E57} = 1.27 \cdot 10^{30} (R/1Mpc)^2, W \qquad (3)$$

If the energy spectrum of cosmic rays is assumed to be $\sim dE/E^3$ down to 60 GeV, then the luminosity should be as large as

$$LCR_{G60} = 10^{39} (R/1Mpc)^2, W \qquad (4)$$

Here $E57$ and $G60$ are indexes for cosmic rays with energies above 57 Eev and 60 Gev correspondently. With $R = 100$ Mpc we obtain that $LCR_{E57} = 10^{34}$ W, and $LCR_{G60} = 10^{43}$ W, so we can immediately exclude acceleration of UHECRs in low power AGNs (e.g. Cen A and M87), low power BL Lacs and starburst galaxies (e.g. M82 and NGC 253) [46].

It should be stressed that the formula (3) gives the luminosity of cosmic rays only with energies $E$ above 57 EeV (near 10 J). Another way to estimate the cosmic ray luminosity is to assume that one cosmic ray particle with the energy above 57 EeV hits each square kilometer on the sphere with a radius $R=78.6$ Mpc every hundred years. Then the power of losses is easily estimated as $1.64 \cdot 10^{35}$ W. Assuming that nearly ~300 objects are in the field of view of the PAO we arrive to the mean estimate of the cosmic ray luminosity $LCR = 5.6 \cdot 10^{32}$ W.

These very simple assumptions allow us to construct the Table 1. The first 4 columns of the table show the number of the Auger events, galactic coordinates and the energy of an EAS. The column 5 displays the celestial objects from [21] which were found to be within ±3.1° from shower arrival direction. Galactic coordinates, a distance $R$ from the Earth and an optical luminosity $L_0$ of the AGN found occupy next four columns. The cosmic ray luminosity *LCR* occupies of the 10-th column and the Roentgen and gamma luminosity *LR* taken from [47, 48, 49] is shown for comparison in the last 11-th column.

Table 1

| 1 | 2 | 3 | 4 | 5 | 6 | 7 | 8 | 9 | 10 | 11 |
|---|---|---|---|---|---|---|---|---|---|---|
| | Showers | | | AGN | | | | | | |
| № | Glong, deg | Glat, deg | E, EeV | Name | Glong, deg | Glat, deg | R, Mpc | $L_0$, $10^{36}$ W | LCR, $10^{32}$ W | LR, $10^{35}$ W |
| 1 | 15.4 | 8.4 | 70 | | | | | | | |
| 2 | 309.2 | 27.6 | 84 | ESO 383-G18 | 312.8 | 28.1 | 54.6 | 3.30 | 3.56 | |
| 3 | 310.4 | 1.7 | 66 | 4U 1344-60 | 309.8 | 1.5 | 54.6 | 0.02 | 3.56 | |
| 4 | 392.3 | -17.0 | 83 | ESO 139-G12 | 332.4 | -14.7 | 74.4 | 5.64 | 6.08 | |
| 5 | 325.6 | 13.0 | 63 | IC 4518A | 326.1 | 14.0 | 67.2 | 0.04 | 5.39 | |
| 6 | 284.4 | -78.6 | 84 | NGC 424 | 283.2 | -78.3 | 46.2 | 1.46 | 2.55 | |
| 7 | 58.8 | -42.4 | 71 | | | | | | | |
| 8 | 307.2 | 14.1 | 58 | NGC 4945<br>ESO 269-G12 | 305.3<br>303.9 | 13.3<br>16.0 | 8.4<br>67.2 | 0.04<br>1.81 | 8.42<br>5.39 | 2.03 |
| 9 | 4.2 | -54.9 | 57 | | | | | | | |
| 10 | 48.8 | -28.7 | 59 | Zw 374.029<br>UGC 11630 | 50.5<br>47.5 | -26.0<br>-25.4 | 54.6<br>50.4 | 0.83<br>5.87 | 3.56<br>3.03 | |
| 11 | 256.3 | -10.3 | 84 | | | | | | | |
| 12 | 196.2 | -54.4 | 78 | SDSS J03349-0548 | 191.7 | -45.7 | 75.6 | 0.18 | 6.82 | |
| 13 | 332.4 | -16.5 | 59 | ESO 139-G12 | 332.4 | -14.7 | 71.4 | 5.64 | 6.08 | |
| 14 | 307.7 | 7.3 | 79 | IC 4200 | 305.8 | 10.8 | 54.6 | 2.69 | 3.56 | |
| 15 | 88.8 | -47.1 | 83 | NGC 7591 | 85.8 | -49.4 | 71.4 | 6.36 | 6.08 | |
| 16 | 189.4 | -45.7 | 69 | SDSS J03302-0532<br>NGC 1358<br>MARK 607 | 190.4<br>190.6<br>186.4 | -46.5<br>-45.6<br>-46.2 | 54.6<br>54.6<br>37.8 | 0.25<br>5.47<br>1.09 | 3.56<br>3.56<br>1.70 | |
| 17 | 308.8 | 17.2 | 69 | NGC 5244 | 311.5 | 16.2 | 33.6 | 1.65 | 1.35 | |
| 18 | 302.8 | 41.8 | 148 | | | | | | | |
| 19 | 63.5 | -40.2 | 58 | Q2207+0122 | 63.0 | -41.8 | 54.6 | 0.03 | 3.56 | |
| 20 | 308.6 | 19.2 | 70 | ESO 323-G77<br>NGC 5128 | 306.0<br>309.5 | 22.4<br>19.4 | 63.0<br>4.2 | 5.18<br>0.04 | 4.74<br>2.10 | 20.12 |
| 21 | 250.6 | 23.8 | 64 | ESO 565-G10<br>NCG 2989 | 253.9<br>253.0 | 21.7<br>26.0 | 63.0<br>54.6 | 4.15<br>1.44 | 4.74<br>3.56 | |
| 22 | 196.2 | -54.4 | 78 | MCG-02.09.040<br>NGC 1204 | 198.2<br>194.2 | -51.1<br>-55.5 | 63.0<br>58.8 | 2.65<br>2.20 | 4.74<br>4.12 | |
| 23 | 318.3 | 5.9 | 64 | | | | | | | |
| 24 | 12.1 | -49.0 | 90 | NGC 7135<br>IC 5135 | 10.1<br>10.0 | -50.6<br>-50.4 | 29.4<br>67.2 | 1.26<br>3.90 | 1.03<br>5.39 | |
| 25 | 338.2 | 54.1 | 71 | NGC 5506 | 339.2 | 53.8 | 29.4 | 0.47 | 1.03 | |
| 26 | 294.9 | 34.5 | 80 | | | | | | | |
| 27 | 234.8 | -7.7 | 69 | | | | | | | |

With the help of this table and the VCV catalog it is possible to make some analysis. The total cosmic ray luminosity *LCR* $_{E57}$ of all sources inside a bin (with a width of 8.4 Mpc) divided by the number $N_i$ of the AGN inside the same bin is shown in Fig. 10. So it is assumed that all AGN inside a bin are capable to accelerate cosmic ray particles but only a few of them have been recorded by chance as sources. In agreement with [31,32] there are no cosmic ray sources in the interval 5 − 25 Mpc where the Virgo cluster is placed. Then a rather strange tendency of increasing luminosity with a distance is displayed. If the AGN are standard sources then we should expect nearly constant luminosity. As some kind of efficiency of the cosmic ray sources the ratio $\beta$ of cosmic ray luminosity to the optical one is approximately equal to $\sim 10^{-4}$. It is not easy to understand the growing efficiency with the distance. Besides, this efficiency is rather high. If we assume that the cosmic ray particles are produced with energy spectrum $dE/E^3$ up to energies ~100 GeV, then a value $\beta$ should be multiplied by a factor of $\sim 10^9$ ! It is much above the optical luminosity. The Fig. 11 displays the normalized number $\alpha$ of cosmic ray sources where $\alpha$ is defined as

$$\alpha = \frac{n_i / N_i}{\sum_{i=1}^{9} n_i / N_i} \quad (5)$$

in case of observed sources (open circles) and

$$\alpha = \frac{(n_i / N_i)/r_i^2}{\sum_{i=1}^{9} (n_i / N_i)/r_i^2} \quad (6)$$

in case of expected sources (full circles).

Here $n_i$ is a number of the cosmic ray sources in the bin and $N_i$ is a number of the AGN in the same bin. A coefficient $(1/r_i)^2$ is due to expected decreasing of cosmic ray luminosity with a distance. A dramatic disagreement is seen. The $\chi^2$ test gives $\chi^2 =$

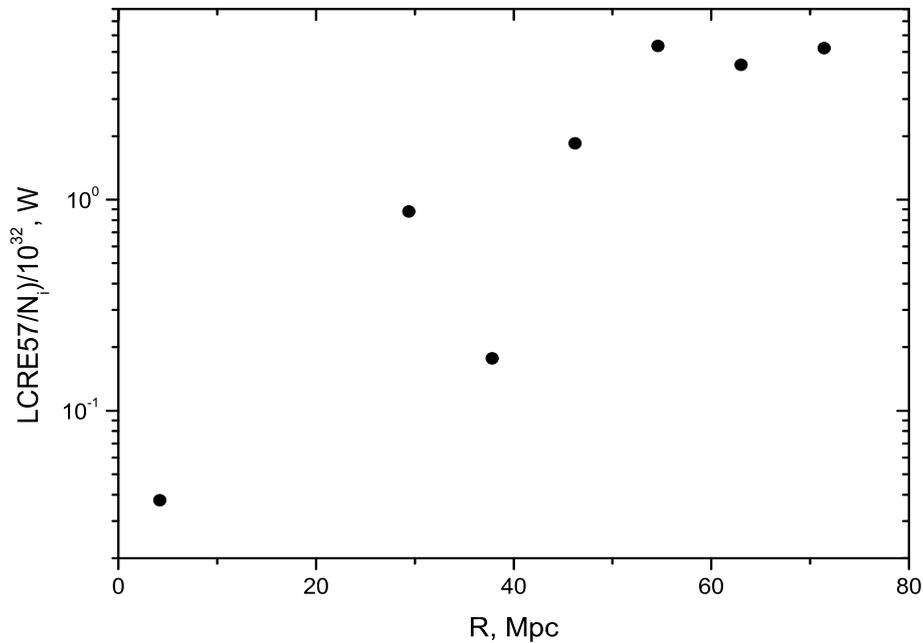

Figure 10. The mean UHECR luminosity $LCR_{E57}$ vs. distance $R$.

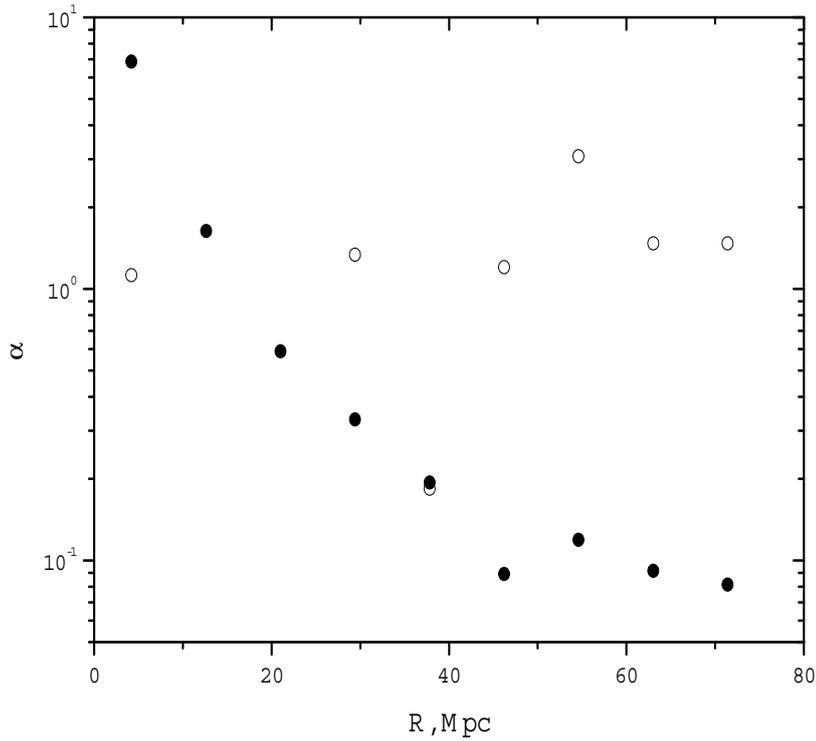

Figure 11. Distribution of normalized numbers of the AGN supposed to be the cosmic ray sources (open circles) and expected numbers (full circles) with distance *R*.

19 per one degree of freedom. Again it is not easy to understand why at larger distances we have approximately the same number of cosmic ray sources as in the first bin. For comparison the total cosmic ray luminosity of our Galaxy is estimated as $5 \cdot 10^{33}$ W [50] that is also should be confronted with the optical luminosity $2 \cdot 10^{37}$ W.

Therefore, the correlation between the CR events observed by the PAO and the local active galaxies may be considered as resulting from a coincidence [51].

## Conclusion

In framework of QGSJET-II and Gheisha-2002d models with codes CORSIKA 6.500, EGS4 and GEANT4 time-dimensional evolution of signals in scintillation detectors used at the Yakutsk array was obtained. Calculated signal time distributions significantly depend on distance from shower axis. At the distances 100, 600, 1000 and 1500 m from shower axis 95% of signal is collected within 100 ns and 1, 2.5 and 4 μs correspondingly. Once again the calculated EAS energy estimation by signal *s*(600) is about 1.7 times lower than used in Yakutsk [14]. It was shown that collection time of signal *s*(600) is about 2 μs and that corresponds to the Yakutsk array parameters [52]. So there are no errors in signal *s*(600) measurements, but problem of its calibration remains. Calculated EAS energy estimation formulae allows to bring in agreement within good accuracy intensity of primary cosmic rays flux measured at the Yakutsk array with data from HiRes and PAO. An approximation of shower forefront by spherical front with radius 17 km was obtained. Estimating energy we suggest that all detector reading should be compared with predictions instead of exploiting only the parameter such as signal *s*(600).

We have to admit that either the nearby AGN emit only cosmic rays particles with energies above ~10 J (~6·10$^{19}$ eV) or their cosmic ray luminosity considerably exceeds the optical one (the AGN are profound cosmic rays accelerators!). Some alternative way is to assume really nearby sources [53, 54, 55, 56, 57]. In this case, the GZK suppression of energy spectrum of cosmic rays is not expected. The most valuable contribution to the problem would be construction of the Northern PAO and the Telescope Array (TA) to claim correlations with any objects on the basis of much more larger statistics.

## Acknowledgements

Authors thank RBFR (grant 07-02-01212) and G.T. Zatsepin LSS (grant 959.2008.2) for support.